\documentclass[prb,twocolumn,showpacs,floatfix]{revtex4}
\usepackage{graphicx}
\usepackage{dcolumn}% Align table columns on decimal point
\usepackage{bm}% bold math
\begin{document}

\title{Numerical simulations of two dimensional magnetic domain patterns}
%Stripes, bubbles, labyrinths and all that}
\author{E. A. Jagla}
\affiliation{The Abdus Salam International Centre for Theoretical Physics\\
Strada Costiera 11, (34014) Trieste, Italy}
\date{\today}

\begin{abstract}

I show that a model for the interaction of magnetic domains that
includes a short range ferromagnetic
and a long range dipolar anti-ferromagnetic interaction reproduces very well many
characteristic features of two-dimensional magnetic domain patterns. In particular
bubble and stripe phases are obtained, along with polygonal and labyrinthine
morphologies. In addition, two puzzling phenomena, namely the so called `memory effect' and the 
`topological melting' observed experimentally are also qualitatively described.
Very similar phenomenology is found in the case in which the model is changed
to be represented by the Swift-Hohenberg equation driven by an external 
orienting field.

\end{abstract}

\maketitle

\section {Introduction}

There is a surprisingly large number of systems that exhibit macroscopic textures arising
from microscopic interactions. \cite{1}
To be concrete, I will take as a case of study that of patterns in magnetic 
systems (magnetic garnets\cite{garnets} or ferrofluids\cite{ff}), but many of the 
conclusions obtained can be directly applied to other systems, 
as for instance, the mixed state of type I superconductors of slab
geometry,\cite{2} and Langmuir monolayers.\cite{langmuir}
The phenomenology of these systems is qualitatively understood 
as appearing from the competition of two effects:
a short range rigidity, and a long range
(dipolar) interaction between the local magnetization
at different spatial positions. 
Calculations suggest \cite{doniach} that the ground state of the system 
consists of (i) a state of uniform magnetization,
(ii) a hexagonal lattice of bubbles
in a background with opposite magnetization, or (iii)
a phase with alternating, parallel stripes of opposite magnetization. The parameter controlling which 
of these three
is actually the ground state is the external magnetic field. However, in experiments, upon 
variation of the external field, different (typically metastable) flux configuration develop that originate
in instabilities of the bubbles or the stripes.
Most noticeable, these metastable configurations include labyrinthine 
phases of interpenetrating domains, and polygonal-like
patterns.\cite{1}

Model Hamiltonians that take into account the two relevant
energy scales have been used to reproduce most of the elemental 
instabilities observed in experiments, in particular:
the elongation and `fingering' instability of
bubbles,\cite{4} and the undulation instability of stripes.\cite{11} 
However, the much richer behavior of the full system, appearing from complex
interaction effects in rather large spatial regions has not been studied in detail with this
kind of models.
In fact, it is not known if these simple models contain all necessary ingredients
to produce realistic magnetization patterns over large spatial scales.

The main motivation of the present work is to present 
large scale simulations using a model Hamiltonian 
to see whether it can account for the full phenomenology and the variety of morphologies
observed.
I claim that the answer is positive.
The simulations are able to reproduce, in particular, two 
phenomena that have been observed in these systems and have remained largely
as puzzles, namely, the so called `memory effect'\cite{molho} of some magnetic patterns, and the `topological
melting'\cite{tm} of an ordered lattice of bubbles.

\section{Details on the model and the numerical technique}

The model I will use is not at all new (see [1], [11] and references therein).
I will consider a scalar field $\phi({\bf r})$ defined over the $x$-$y$ plane. 
This variable will represent the magnetization in the system, that in experiments is typically
constrained (because of structural properties) to point perpendicularly to the $x$-$y$ plane. 
Then, experimentally, the magnetization $\phi$
has a preference to take two different values, that
without loose of generality I will assume to be $\pm 1$. It 
will be convenient for the simulations to consider $\phi$ as a continuum variable
and include in the Hamiltonian a local term $H_l$ that favors the values $\phi=\pm 1$.
This term will be of the form 
\begin{equation}
H_l=\alpha_0 \int d{\bf r} \left (-\frac{\phi({\bf r})^2}2+\frac{\phi({\bf r})^4}4\right) -h_0\int d{\bf r}\phi({\bf r})
\end{equation}
and represents the simplest continuum field description of an Ising variable. Note that
a term describing the effect of an external magnetic field $h_0$ has been already included.

The other terms that will be included in the Hamiltonian are 
the following. First, there is a rigidity term
$H_{rig}$ of the form
\begin{equation}
H_{rig}=\beta_0 \int d{\bf r} \frac {\left |\nabla \phi({\bf r}) \right|^2}2
\label{atrac}
\end{equation}
This term (with positive $\beta_0$) discourages spatial variations of $\phi$, and
can be called `attractive', in the sense that two regions with a value of $\phi$ of plus (or
minus) one, in a background of the opposite sign, tend to merge into a single one 
to reduce the value of this term (in fact, in a description in terms of an Ising variable
on a lattice, this term
maps onto a ferromagnetic interaction between nearest neighbor sites).
The fact that our fundamental variable $\phi$ is continuous rather than discrete, and the existence of the
gradient term, imply in particular the existence of a natural width 
(of the order of $\sqrt{\beta_0/\alpha_0}$) for the interface between domains with
positive and negative magnetization. Choosing the parameters
in such a way that this width is a few times the discretization distance in the simulation, 
allows to obtain a smooth
interface between domains, which turns out to be very weakly pinned by the underlying numerical mesh, 
and whose energy is almost independent of its spatial
orientation. These two facts are crucial for a realistic simulation, and cannot be easily achieved
using a Ising variable that takes only two values (see for instances the attempts in [12]).\cite{pfm}

Secondly, there is a term $H_{dip}$ that models the dipolar interactions, of the form
\begin{equation}
H_{dip}=\gamma_0 \int d{\bf r}d{\bf r}' \phi({\bf r})\phi({\bf r}') 
G({\bf r},{\bf r}')
\label{repul}
\end{equation}
where $G({\bf r},{\bf r}') \sim 1/|{\bf r}-{\bf r}'|^3$ at long distances.
At short distances however, the $r^{-3}$ behavior has to be cut off
to avoid divergences (in experiments, the cut off distance is given roughly by the
thickness of the film). 
However, we can see that the way in which the cut off is
done is not crucial for the results. In fact, we will take advantage of the fact
that the two terms (\ref{atrac}) and (\ref{repul}) can be compactly written in
Fourier space as

\begin{equation}
H_{rig}+H_{dip}= \sum_{\bf k} |\phi({\bf k})|^2 (\beta_0 k^2 +\gamma_0 G_{\bf k})
\end{equation}
where $G_{\bf k}$ is the Fourier transform of $G({\bf r},0)$.
Thus, it is the combination $(\beta_0 k^2 +\gamma_0 G_{\bf k})$ that will mostly
determine the behavior of the system. Note that the short distance behavior of
$G$ in real space is masked in Fourier space at large $k$ by the $k^2$ term, and then is
irrelevant. On the other hand, the $r^{-3}$ behavior at long distances transforms
into a $k$ dependence of the form 
\begin{equation}
G_{{\bf k}\rightarrow 0}=a_0-a_1|k|.
\label{gdek}
\end{equation}
The 
constants can be exactly evaluated to be 
\begin{eqnarray}
a_0&=&2\pi \int_0^{\infty} rd{r} G(r)\label{a0}\\
a_1&=&2\pi
\end{eqnarray}
The finite value $a_0$ of $G_{\bf k}$ at $k=0$ reflects the
fact that the interaction in real space is integrable (in spite of
being sometimes called `long range'). Also note that $a_1$ is independent of the short
distance behavior of $G(r)$.
The main features of the interaction in Fourier space are the maximum with finite derivative at
$k\rightarrow 0$, and 
the minimum at a finite wave number
$k_{min}\sim {\gamma_0/\beta_0}$. This minimum exists for any non-zero $\gamma_0$, indicating that
the effect of the dipolar interactions on large distances can never be neglected.

We have defined the energy function of the system, and now the dynamics 
has to be introduced.
Since in magnetic systems the magnetization is a non-conserved order parameter, I will
use the Allen-Cahn\cite{chaikin} dynamical equations, namely

\begin{eqnarray}
&&\frac{\partial\phi ({\bf r})}{\partial t}=-\lambda \frac{\delta (H_l+H_{rig}+H_{dip})}{\delta \phi ({\bf
r})}=\nonumber\\
&&=-\lambda \left( \alpha_0 (-\phi+\phi^3)-h_0 -\beta_0 \Delta \phi 
+\gamma_0\int d{\bf r}' \phi({\bf r}') G{(|{\bf r}-{\bf r}'|)}\right)
\label{ecreal}
\end{eqnarray}
that represents an overdamped dynamics in which the system reduces its energy 
by a steepest descendant evolution.
To efficiently implement these equations on the computer, and in order to avoid the direct evaluation 
of the integral in the last term of (\ref{ecreal}), a pseudo-spectral method\cite{ps} is used.
I write the previous equation in Fourier space,
namely
\begin{equation}
\frac{\partial\phi_{\bf k}}{ \partial t}=-\lambda\left[
\left .\alpha_0 (-\phi+\phi^3)\right |_{\bf k}- h_0\delta({\bf k}) +(\beta_0 k^2  +\gamma_0 G_{\bf k})\phi_{\bf k} \right]
\label{ecfourier}
\end{equation}
In this way, the last term is now algebraic. The complication has been translated to the first term, that
involves the evaluation of the Fourier transform of $\phi^3$. However, this
can be done very efficiently by the use of 
standard fast-Fourier-transform techniques.

In the simulations below, the function $G$ is defined in real space to be
$G({\bf r},{\bf r}')=1/|{\bf r}-{\bf r}'|^3$ 
for any two points of the numerical mesh such that ${\bf r}\ne{\bf r}'$, whereas $G({\bf r},{\bf r})\equiv 0$. Then the cut off
distance is the lattice discretization. The Fourier transform of this expression on the square lattice
gives for the relevant terms of $G_{\bf k}$ the form of Eq. (\ref{gdek})
with $a_0\simeq 9.05$, $a_1=2\pi$. 
Once the value of $a_0$ is fixed (and since the value of $a_1$ is universal),
there are four independent coefficients in (\ref{ecfourier}).
Two of them can be fixed by rescaling the spatial and temporal
coordinates. In fact, if we define a new field 
$\tilde\phi(r,t)\equiv A^{-1}\phi(r/C,t/B)$ (and then $\tilde\phi_{\bf k}(t)\equiv A^{-1}\phi_{C{\bf k}}(t/B)$),
and in case we choose $A$  to be 
\begin{equation}
A=\sqrt{1+\frac{a_0\gamma_0(C-1)}{\alpha_0}},
\end{equation}
the new field satisfies equations of motion that in Fourier space can be written as (the tilde in the new field
has been eliminated for simplicity):

\begin{equation}
\frac{\partial\phi_{\bf k}}{ \partial t}=
\left .\alpha (\phi-\phi^3)\right |_{\bf k}+ h\delta({\bf k}) -(\beta k^2  +\gamma G_{\bf k})\phi_{\bf k},
\label{ecfourier2}
\end{equation}
with
\begin{eqnarray}
\alpha&&\equiv\frac{\lambda\alpha_0A^2}{B}\\
h&&\equiv\frac{\lambda h_0}{AB}\\
\beta&&\equiv\frac{\lambda\beta_0C^2}{B}\\
\gamma&&\equiv\frac{\lambda\gamma_0C}{B}\label{gama}
\end{eqnarray}
and where $G_{\bf k}$ is (up to the linear terms that are relevant for our
analysis) the same function as before, namely $G_{\bf
k}=a_0-a_1|k|$ with the same $a_0$ and $a_1$.
This renormalization can be used to fix two parameters in the new non-dimensional equations 
(\ref{ecfourier2}). In the simulation presented below I have fixed $\beta=2.0$, $\gamma=0.19$ and took the
spatial discretization to be the unit of length (this choice
was convenient when implementing the equations on the numerical mesh, and have no other particular meaning).
Therefore, we see that in addition to the external control parameter $h$, a single internal
control parameter $\alpha$ remains. This parameter 
regulates the possibility of the field $\phi$ to take values others
than the most convenient ones, namely $\phi=\pm 1$.
We will see below the different morphologies that appear for different values of $\alpha$.
From now on I will always refer to the non-dimensional form (\ref{ecfourier2}) of the equations of motion.

Starting from an arbitrary initial condition, Eq. (\ref{ecfourier2}) describes an evolution in which the
total energy of the system $H_l+H_{rig}+H_{dip}$ is steadily reduced until it reaches a minimum, in which
$\partial\phi_{\bf k}/{\partial t}$ is identically zero. We will see that typically the true minimum of
the system is not reached, but instead one of many possible metastable states is obtained. 
The simulations presented below were done on a 512 $\times$ 512 mesh using periodic boundary conditions. 
The time-integration of the equations is done using a semi-implicit first order method, in which the
$k^2$ term in Eq. (\ref{ecfourier2}) is evaluated in the new time value. Concretely, I use an iteration scheme
based on the following discretized form of (\ref{ecfourier2})
\begin{equation}
\frac{\phi_{\bf k}^{t+\delta t}-\phi_{\bf k}^{t}}{\delta t}=\left .\alpha (\phi-\phi^3)\right |_{\bf k}^t
+ h\delta({\bf k})  -\gamma G_{\bf k}\phi_{\bf k}^t-\beta k^2\phi_{\bf k}^{t+\delta t}.
\end{equation}
This treatment of the diffusive term is standard to improve the stability of the algorithm.\cite{numrec}
In all cases below the time interval used is $\delta t=0.5$.
%To obtain the results in each of the
%figures presented below (implying typically a full sweep of the external field) 
%about 10 hours of CPU time on a desktop computer were needed.

\section{Results}

The initial condition for the variable $\phi$ is taken to be locally random in the
interval $-1<\phi<1$, and the system is evolved during an equilibration time $t_{start}$ 
in the presence of a fixed applied external field $h_{start}$.
If $h_{start}$ is too large, the configuration obtained turns out to be a state of
uniform magnetization. However, for lower $h_{start}$, a structure of bubbles of the
minority phase (with magnetization anti-parallel to the field) 
within a background of the opposite magnetization may be favored. 

\begin{figure}
\includegraphics[width=8.5cm,clip=true]{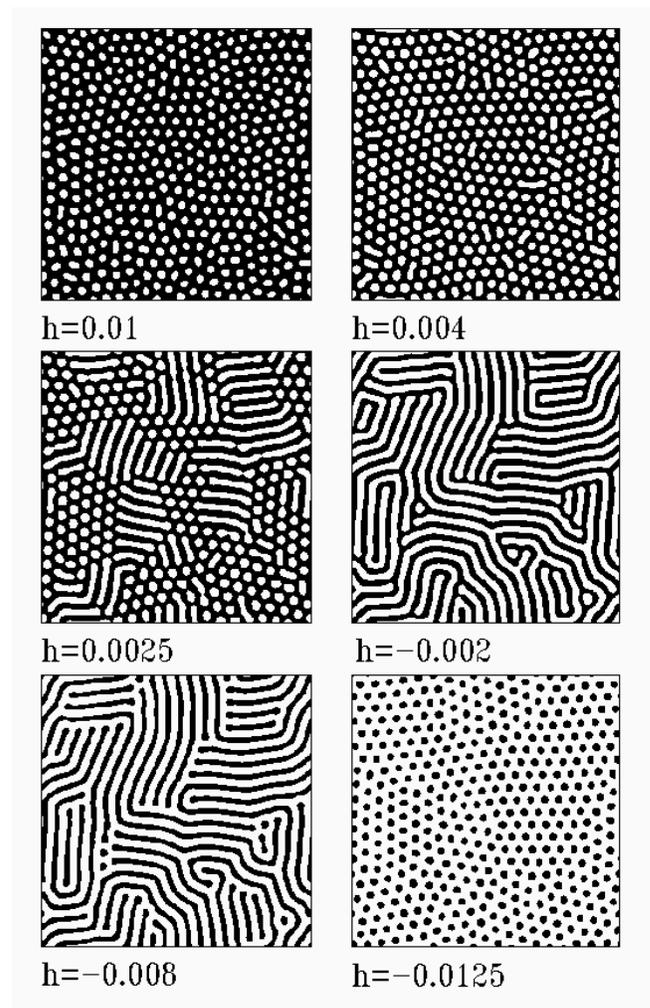}
\caption{\label{f1}
Evolution of the magnetization distribution upon reduction of the magnetic field $h$, 
for $\alpha=1.6$. Other parameters are: $t_{start}=3000$, $h_{start}=0.01$, $dh/dt=-5\times 10^{-7}$ (see text). Here
and in the following figures black (white) indicates regions with positive (negative) magnetization, 
all parameters are in the non-dimensional form corresponding to Eq. 
({\protect\ref {ecfourier2}}), and system size is $512\times 512$.}
\end{figure}

After the time $t_{start}$, the field is decreased as a function of time with a finite rate $dh/dt$. This value
is taken to be as small as possible (within reasonable computing time) in order that the
field change can be considered to be adiabatic (we will see that this cannot always be guaranteed due to the
existence in some cases of field driven instabilities).
During the evolution, different morphologies are observed for different values of $\alpha$
in Eq. (\ref{ecfourier2}), which will be described now.

{\em Almost reversible interconversion of bubbles and stripes.} 
For $\alpha=1.6$, the result obtained is shown in Fig. \ref{f1}. 
Starting from the initial bubble phase, upon reduction of the field $h$, neighbor 
bubbles coalesce, forming a striped pattern. 
When the field becomes negative, 
the stripes
destabilize, and separate in a chain of bubbles, which have opposite 
magnetization with respect to the original ones. The sequence of bubble and stripe patterns
is found to be reversible upon cycling of the field. There 
is however a noticeable hysteresis in the field value at which the bubble-stripe interconversion
occurs. This is just the consequence of the transition between bubbles and stripes being first
order. \cite{doniach}

\begin{figure}
\includegraphics[width=8.5cm,clip=true]{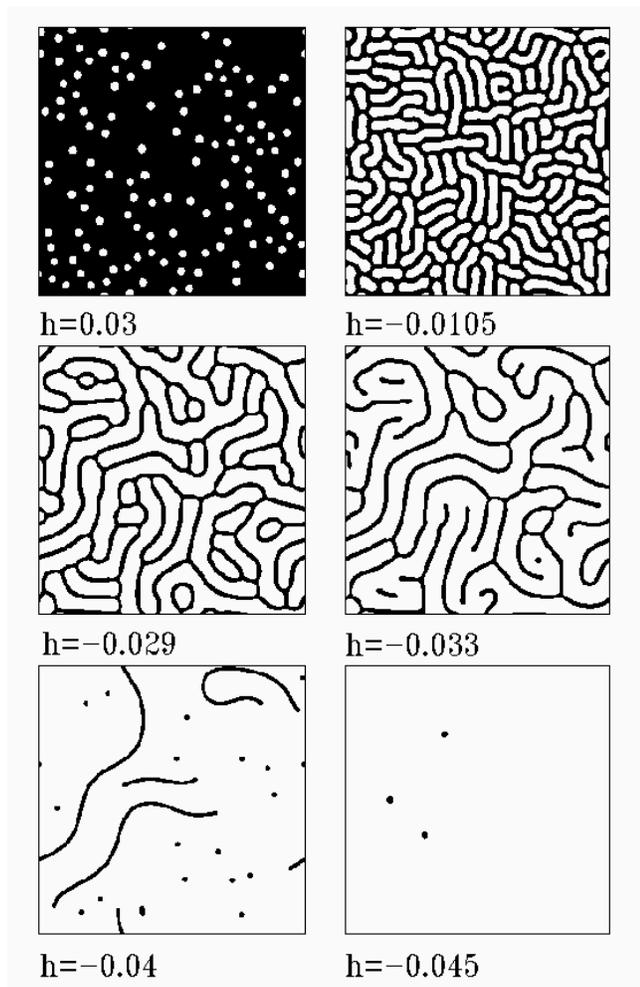}
\caption{\label{f2} 
Same as Fig. {\protect \ref{f1}} for $\alpha=1.8$ ($t_{start}=1500$, $h_{start}=0.03$, 
$dh/dt=-3\times 10^{-6}$).}
\end{figure}

%Actually, the present case is very well known to occur, for instance, in systems with charge
%interactions (where $G(r)\sim r^-1$). In that case however, the bubble-stripe interconversion
%is the {\em only} phenomenology observed, whereas we will see immediately that here the behavior is
%much richer.

{\em From bubbles to rather isolated and wandering stripes.} 
For a slightly larger value of $\alpha$, namely $\alpha=1.8$ (Fig. \ref{f2}), the bubbles 
may become unstable and elongate individually, without merging with
their neighbors at the beginning.
When they finally merge 
(for $h\lesssim -0.025$),
regions with positive magnetization generate wavy stripes of well defined thickness. 
Contrary to the previous case, these regions
do not separate into `beads' when the field is made more negative, but eventually retract back
to a single spot of positive magnetization 
that eventually disappears. 

\begin{figure}
\includegraphics[width=8.5cm,clip=true]{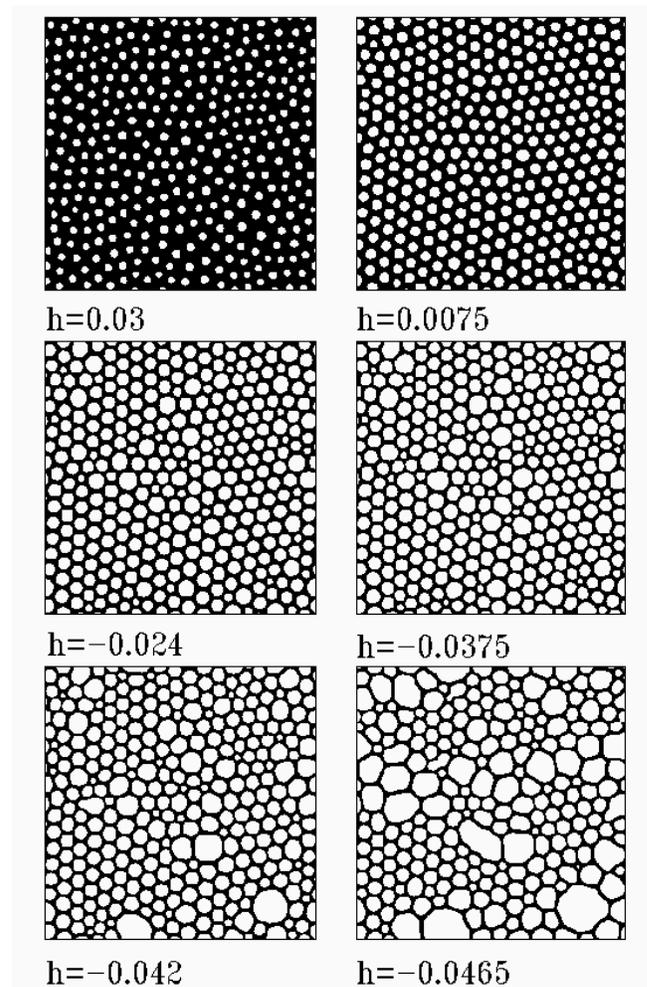}
\caption{\label{f5}
Same as Fig. {\protect \ref{f1}} for $\alpha=2.2$ ($t_{start}=4500$, $h_{start}=0.03$, $dh/dt=-1\times 10^{-6}$).}
\end{figure}

\begin{figure}
\includegraphics[width=8.5cm,clip=true]{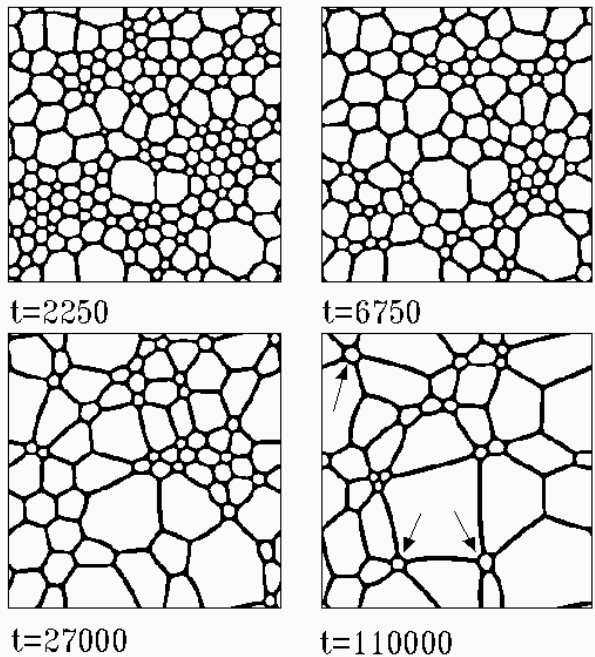}
\caption{\label{f5b}
The final configuration if Fig. {\protect \ref{f5}} evolved at constant field $h=-0.0465$ as a function of time, 
as indicated ($t=0$ corresponds to the last panel in Fig. {\protect\ref{f5}}). Arrows in the last panel
highlight some small pentagonal bubbles, a
structure that appears ubiquitous both in experiments an in the simulations. 
}
\end{figure}

{\em Collapse of the bubbles to a polygonal pattern.} 
For larger values of $\alpha$, the bubbles are seen to remain (meta-) stable down to a field where they start
to merge with their neighbors, but now in a sort of two dimensional way,
as seen in Fig. \ref{f5} for $\alpha=2.2$. This has to be compared with the 
previous case where the initial collapse of bubbles
was mainly one-dimensional, generating stripes (see the cases $h=-0.029$ and $h=-0.033$ in Fig. \ref{f2}).
In the present case, the collapse of neighbor bubbles seems to occur as a cascade process, where some 
initial coalescences
trigger the full transition of the lattice. 
In fact, in Fig. \ref{f5b} the field was kept constant at the value $h=-0.0465$ (corresponding to the last panel
in Fig. \ref{f5}), and 
the evolution was followed as a function of time. A coarsening process is occurring here. 
Actually, the
last pattern in Fig. \ref{f5b} is not totally relaxed yet.
Incidentally note in the last 
panel of Fig. \ref{f5b} the
existence of small pentagonal bubbles, highlighted by the arrows. This structure
has been observed experimentally to be ubiquitous, and very stable.\cite{tm}

\begin{figure}
\includegraphics[width=8.5cm,clip=true]{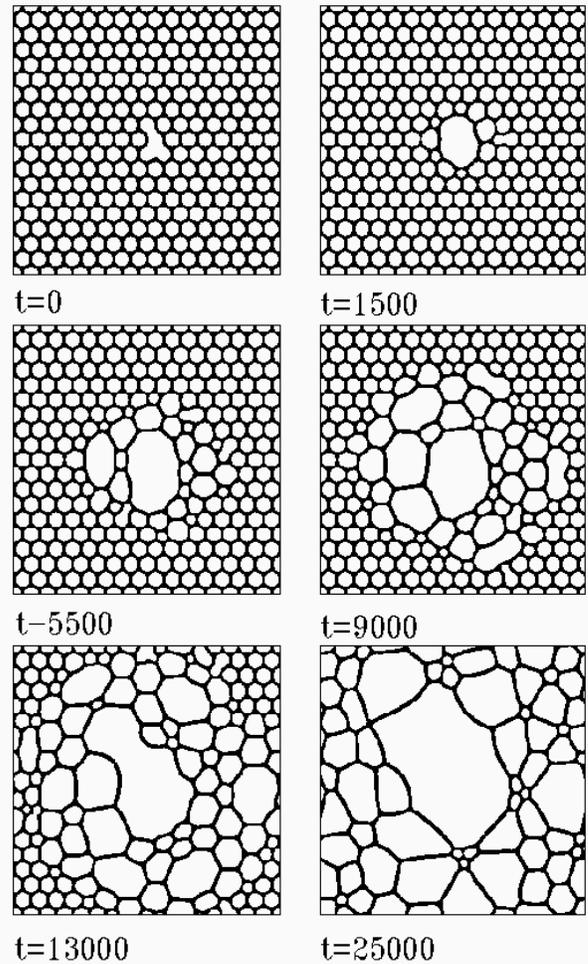}
\caption{\label{tm}
Topological melting of an order array of bubbles for $\alpha=2.2$, upon the {\em ad hoc}
inclusion of a defect in the middle of the sample.
The evolution occurs at a fixed value 
of the field $h=-0.05$, as a function of the simulation time, as indicated. }
\end{figure}

This case suggests the
following interesting result:
if a perfect original pattern of bubbles is constructed by
hand, it can remain stable for values of the field at which the disordered bubble system would 
have already collapsed. Now, if in this ordered, metastable structure,
a defect is introduced, it can completely disorder the lattice.
In fact, we see in Fig. \ref{tm} how the presence of the defect produces a sequence of instabilities 
that destroy many of the walls between neighbor bubbles, generating a rather well defined disordering
front that leaves behind a disordered structure with much lower magnetization.
This effect has been experimentally observed and called
{\em topological melting}\cite{tm} of the bubble lattice. It has been observed to occur (although in 
a less dramatic form) also for systems in which the long range interaction is of Coulomb type\cite{muratov}.

\begin{figure}
\includegraphics[width=8.5cm,clip=true]{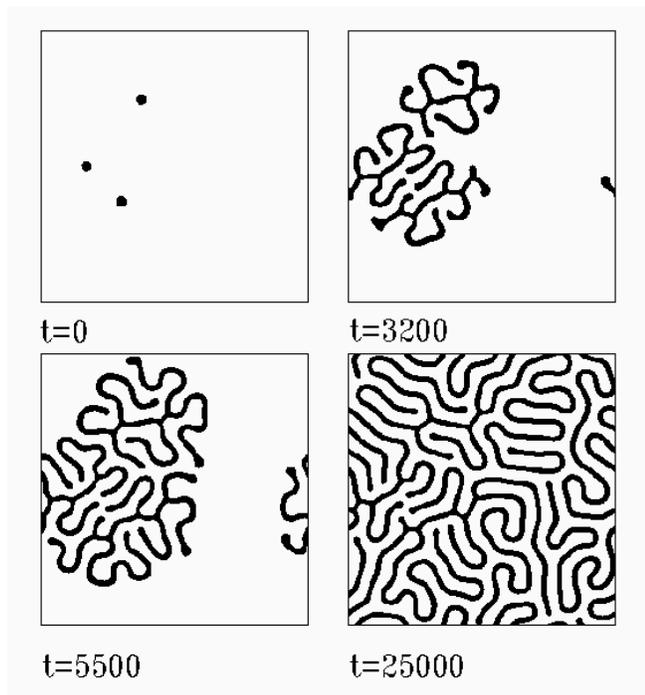}
\caption{\label{f3}
Time evolution of the pattern shown in the first panel upon  the application of a constant field 
($h=-0.018$) slightly beyond
the critical field at which those bubbles destabilize. Times of the snapshots are indicated.}
\end{figure}

\begin{figure}
\includegraphics[width=8.5cm,clip=true]{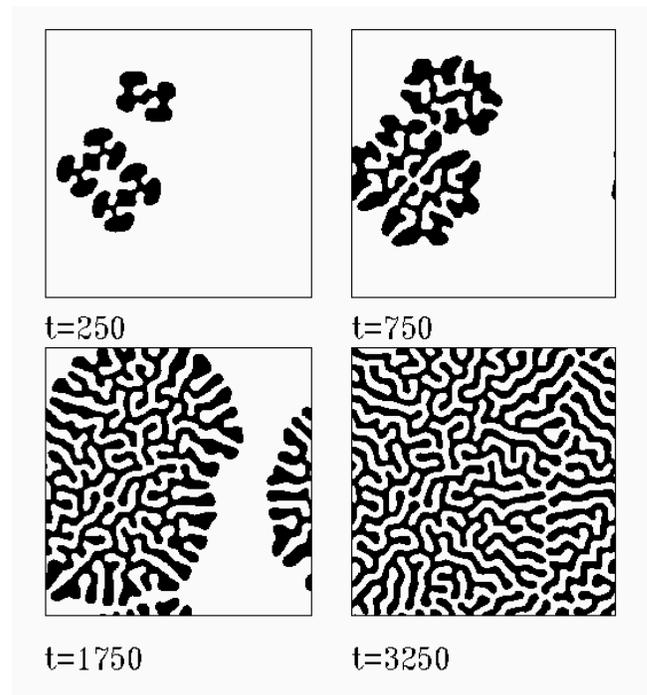}
\caption{\label{f4}
Same as Fig. {\protect \ref{f3}} for $h=0$, i.e., here the system is brought
deeply inside the instability region. Note the much larger amount of stripe branching 
in the final (stable) pattern, and the shorter time scale as 
compared with the previous figure.}
\end{figure}

{\em Labyrinthine patterns and the memory effect.} 
If from the last panels in Figs. \ref{f2} or \ref{f5b} the field is slowly switched off, interesting results are
obtained. In the case in which we start from the configuration of the last panel in Fig. \ref{f2}, 
which contains three
(meta-) stable spots of positive magnetization, they remain stable
(increasing only slightly in size) up to $h\sim -0.02$. At this field an instability occurs, 
the bubbles becoming unstable. If we maintain the field fixed at a value slightly lower (in absolute value)
than the instability value,
we obtain the results 
presented in Fig. \ref{f3}, which shows snapshots as a function of time,
for a fixed value of the field $h=-0.018$. The bubbles elongate and successively branch, forming 
labyrinthine patterns that invade the whole sample. 
On the other hand, if the field that we apply is much beyond the instability value (Fig. \ref{f4})
the evolution is
more rapid, and with a larger degree of branching of the magnetic domains.
Note the difference in the degree of branching in the final patterns of Figs. \ref{f3} and \ref{f4}.
The grater tendency to branching when the applied field is more and more beyond the instability value is
well known experimentally and theoretically\cite{4}. This kind of instability is
also similar to that 
observed is some reaction-diffusion systems\cite{otrag}.

If we reduce the absolute value of the field from a configuration 
in which stripes are already present, we observe an undulation transition\cite{11} at a field with 
larger absolute
value than before. But contrary to what happened in Figs \ref{f3} and \ref{f4}, if the field is changed slowly
the system evolves smoothly (no instability appears), and stripes do not branch. 
Positive magnetization regions invade the system through wandering of
the stripes, but new branches do not appear or are very rare. 

\begin{figure}
\includegraphics[width=8.5cm,clip=true]{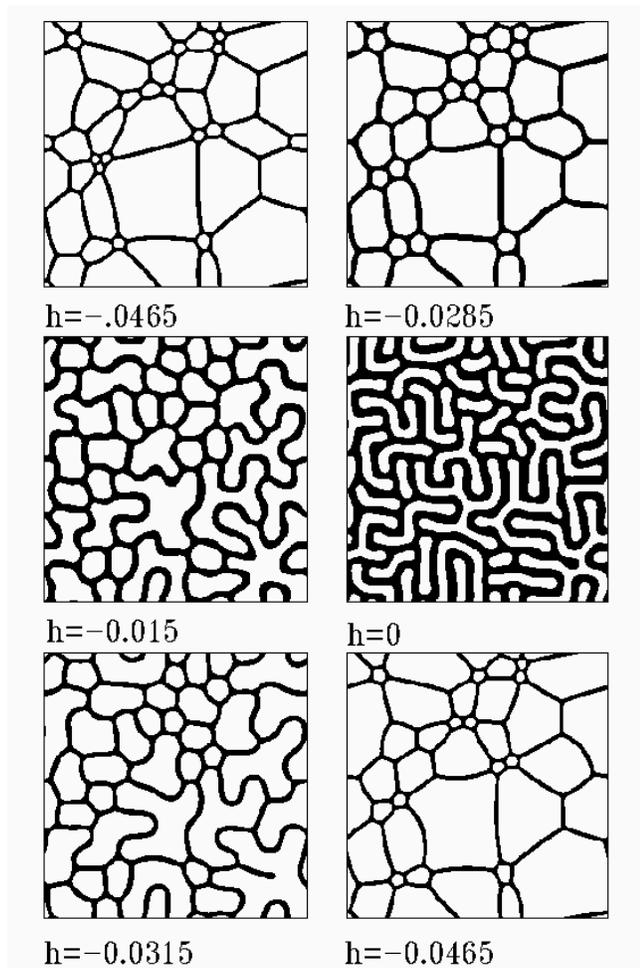}
\caption{\label{memory}
Reducing the field from the final configuration in Fig. {\protect \ref{f5b}} down to $h=0$ and back
to its original value ($|dh/dt|=1\times 10^{-6}$). Note the `memory' of the pattern as comparing first and last panels.}
\end{figure}

In particular, in the case in which we reduce the field starting from the last configuration 
in Fig. \ref{f5b} in which stripes
are abundant as walls between polygons, the
undulation occurs mildly, with almost no breaking or reconnection of the cell walls, and 
then the final labyrinthine pattern
at $h=0$ is topologically equivalent to the original one. This is shown in Fig. \ref{memory}. 
A nice consequence of this, is that when the field
is switched on again
(last panels in Fig. \ref{memory}), 
the original pattern is almost recovered.
This effect, called the {\em memory effect}\cite{molho} has been observed 
experimentally, and the typical evolution of the
patterns over many cycles of the field has been analyzed. We are seen that this effect is contained
in the simple model Hamiltonian we are using.

\section{Discussion and Conclusions}

Summarizing, in the previous Section I have shown how the model equations (\ref{ecfourier2}) 
can be efficiently simulated in systems of reasonably large size. 
In this way, we have seen emerging most of the
phenomenology of two dimensional magnetic patterns and other similar systems.
The success of the present numerical simulations are due to a combination of reasons, mainly: 
the use of a continuum
variable instead of a discrete one to obtain smooth domain walls between regions with opposite
values of $\phi$, and 
the use of pseudo-spectral techniques to evaluate efficiently the `long-range' dipolar force.
These facts combine to allow a realistic simulation of domain patterns that show many 
of the features observed
in experimental realizations. In particular, the {\em memory effect}\cite{molho} and the 
{\em topological melting}\cite{tm} of the system are very well reproduced. 

I want to emphasize that in all cases I have studied, 
the evaluation of the total energy of the system is compatible with the fact that the only
patterns truly corresponding to the ground state of the system are: 
(i) a pattern with uniform magnetization if the field is strong enough, 
(ii) a regular bubble phase for intermediate fields, 
and (iii) a regular stripe phase for low (including zero) field. 
Although this is not a demonstration that they are the only possible ground states,
it points in this direction, and it is in agreement with the results of theoretical studies.\cite{doniach}
The other patterns observed (labyrinthine, polygonal, etc.) are seen to be 
metastable, and they are originated in the particular cycling 
of the field (and in the initial conditions)
to which the sample is subjected. A recent experimental study\cite{japon} 
has shown in fact how the labyrinthine 
patterns converge to parallel stripes upon relaxation.

Very different morphologies have been observed when the parameter $\alpha$ in Eq. (\ref{ecfourier2}) is changed.
Figs. \ref{f5}, \ref{f5b}, and \ref{memory} (corresponding to the largest values of $\alpha$)
compare very well with the patterns observed in magnetic garnets and
ferrofluids (see [1], [10], and [11]). 
The results for lower $\alpha$ (in particular, Figs.
\ref{f1} and \ref{f2}) are more akin to Langmuir
monolayers\cite{langmuir} and flux structures in type I superconductors.\cite{2}
This suggests that in real systems the possibility of the order parameter to take values different than
the two preferred ones can  influence noticeable the physical properties.

I want to mention here that the present model can be also efficiently used to study the effect of quenched
disorder in the system, and the effect of thermal fluctuations. Preliminar results indicate that the model
generates hysteresis curves and magnetization patterns that, 
as a function of the 
amount of disorder, compare very well with
experimental ones\cite{sorensen}. These results will be published
separately.

We have seen that in the present model the dynamics is controlled by an interaction function 
in $k$ space that has a maximum with finite derivative at $k\rightarrow 0$ and a minimum at a finite $k_{min}$ value.
It is worth comparing this case with respect to other possibilities. One is the case in which
the field $\phi$ is considered to be charged, instead of carrying a dipole. 
Two cases can be considered. One is that of
true three dimensional charges ($G(r)\sim r^{-1}$) and the other is the case of two dimensional charges 
($G(r)\sim -\ln(r)$). 
In both cases, the interaction in $k$ space gets a
divergence at low $k$. This model has been studied in detail in
[17] (see the references in there for realizations of this case).     %\cite{muratov}
There, instabilities of a single bubble have been found which are similar to those I find
in the dipolar system. It remains to be seen if the other effects
described here are also present in Coulombic systems.

Another case to compare with is that of interactions decaying in real space more
rapidly than $r^{-3}$. In this case, a $k$ space interaction with a quadratic maximum 
at $k=0$ is obtained. If this maximum dominates over the quadratic minimum coming from 
the $\Delta \phi$ term in Eq. (\ref{ecreal}), then the effective interaction has a quadratic 
maximum at the origin and a minimum at some finite  $k_{min}$. This case corresponds 
qualitatively to the interaction considered in
the Swift-Hohenberg equation.\cite{sh} For this interaction, 
and controlling the same parameter $\alpha$ as I did here, I have
obtained basically all the effects and morphologies 
described in the previous section. On one hand this tells
that the singularity at $k=0$ of the dipolar interaction is not crucial 
in obtaining these effects, a quadratic maximum
suffices. On the other hand it is a bit surprising that
in the wide literature related to the Swift-Hohenberg equation these effects 
have not been described previously. This might be due to the fact that 
the Swift-Hohenberg equation is usually considered in the absence of a
`magnetic-field-like' term that favors one of the two orientations, and this term 
is crucial to obtain the metastable patterns. It is then likely that the much studied 
relaxation to equilibrium properties of the patterns seen in the Swift-Hohenberg 
equation and the coarsening properties of the magnetic patterns (studied
for instance in [9]) %\cite{molho}) 
can be put under the same framework.
I hope the present work encourages some 
studies in this direction.

\section{Acknowledgments}

I thank J. R. Iglesias for useful comments on an early version of the manuscript.

\newpage

\end{document}